\documentclass[10pt,journal]{IEEEtran}

\usepackage{cite}
\usepackage[caption=false]{subfig}
\usepackage{graphicx}
\usepackage{amsmath, amssymb}
\usepackage{booktabs}
\usepackage{tabularx}

\usepackage{array}
\usepackage[hyphens]{url}

\usepackage{color, colortbl}
\definecolor{Gray}{gray}{0.9}
\definecolor{pink}{rgb}{0.858, 0.188, 0.478}

\usepackage{bbold}
\def\xp {\tilde{\mathbf{x}}}

\def\bp {\tilde{b}}
\def\xr {\mathbf{x}}
\def\D {\mathbf{D}}
\def\A {\mathbf{A}}
\def\T {\mathbf{T}}

\begin{document}

\title{Overview and Evaluation of Sound Event Localization and Detection in DCASE 2019}

\author{Archontis Politis, 
        Annamaria Mesaros,
        Sharath Adavanne,
        Toni Heittola,
        Tuomas Virtanen
\thanks{This work received funding from the European Research Council under the ERC Grant Agreement 637422 \mbox{EVERYSOUND}.
}%
\thanks{A. Politis, A. Mesaros, S.Adavanne, T. Heittola and T. Virtanen are with the Faculty of Information Technology and Communication Sciences, Tampere University, Finland, e-mail: \{archontis.politis, annamaria.mesaros, tuomas.virtanen\}@tuni.fi}
}

\maketitle

\begin{abstract} 
Sound event localization and detection is a novel area of research that emerged from the combined interest of analyzing the acoustic scene in terms of the spatial and temporal activity of sounds of interest. This paper presents an overview of the first international evaluation on sound event localization and detection, organized as a task of the DCASE 2019 Challenge. A large-scale realistic dataset of spatialized sound events was generated for the challenge, to be used for training of learning-based approaches, and for evaluation of the submissions in an unlabeled subset.
The overview presents in detail how the systems were evaluated and ranked and the characteristics of the best-performing systems. Common strategies in terms of input features, model architectures, training approaches, exploitation of prior knowledge, and data augmentation are discussed. Since ranking in the challenge was based on individually evaluating localization and event classification performance, part of the overview focuses on presenting metrics for the joint measurement of the two, together with a reevaluation of submissions using these new metrics. 
The new analysis reveals submissions that performed better on the joint task of detecting the correct type of event close to its original location than some of the submissions that were ranked higher in the challenge. Consequently, ranking of submissions which performed strongly when evaluated separately on detection or localization, but not jointly on both, was affected negatively.

\end{abstract}


\section{Introduction}
Recognition of classes of sound events in an audio recording and identification of their occurrences in time is a currently active topic of research, popularized as sound event detection (SED), with a wide range of applications \cite{Mesaros2019_TASLP}. While SED can reveal a lot about the recording environment, the spatial locations of events can bring valuable information for many applications. On the other hand, sound source localization is a classic multichannel signal processing task, based on sound propagation properties and signal relationships between channels, without considering the type of sound characterizing the sound source. A sound event localization and detection (SELD) system aims to a more complete spatiotemporal characterization of the acoustic scene by bringing SED and source localization together. 
The spatial dimension makes SELD suitable for a wide range of machine listening tasks, such as inference on the type of environment \cite{barchiesi2015acoustic}, robotic simultaneous localization and mapping \cite{evers2018acoustic}, navigation without visual input or with occluded targets, tracking of sound sources of interest \cite{mack2020signal}, and audio surveillance \cite{valenzise2007scream}. Additionally, it can aid human-machine interaction, scene-information visualization systems, scene-based deployment of services, and assisted-hearing devices, among others.

The SELD task was included for the first time in the Detection and Classification of Acoustic Scenes and Events (DCASE) Challenge of 2019\footnote{\url{http://dcase.community/challenge2019/}}. In addition to the related studies that aim at detecting and localizing multiple speakers (see e.g. \cite{may2012binaural}), only a handful of approaches could be found in the literature up to that point \cite{valenzise2007scream, butko2011two, chakraborty2014sound, lopatka2016detection, grobler2017sound, hirvonen2015classification, adavanne2018sound}. Earlier studies were treating the two problems of detection and localization separately, without trying to associate source positions and events. In those works, Gaussian mixture models (GMMs) \cite{valenzise2007scream}, hidden Markov models (HMMs) \cite{butko2011two}, or support vector machines \cite{lopatka2016detection} were used for detection, while localization relied on classic array processing approaches such as time difference of arrival (TDOA) \cite{valenzise2007scream}, steered response power \cite{butko2011two}, or acoustic intensity vector analysis \cite{lopatka2016detection}. An early attempt in joining estimates from the two problems was presented in \cite{chakraborty2014sound}, where beamforming outputs from distributed arrays along with an HMM-GMM classifier are used to build a maximum-a-posteriori criterion on the most probable position in a room of a certain class. 

During the last decade, deep neural networks (DNNs) have become the most established method on SED, offering ample modeling flexibility and surpassing traditional machine learning methods when trained with adequate data \cite{Mesaros2018_TASLP}. Recently, DNNs have also been explored for machine learning-based source localization \cite{adavanne2018direction, perotin2019crnn, chakrabarty2019multi} with promising results. Hence, DNNs seem like a good candidate for joint modeling of localization and detection in the SELD task. The first works we are aware of that use this approach are~\cite{hirvonen2015classification } and~\cite{adavanne2018sound}. Hirvonen ~\cite{hirvonen2015classification} proposed to set joint modeling as a multilabel-multiclass classification problem, mapping two event classes to eight discrete angles in azimuth. A convolutional neural network (CNN) was trained to infer probabilities of each sound class at each position, after which a predefined threshold was used to decide the final class presence and location. Adavanne et al. ~\cite{adavanne2018sound} proposed as an alternative a regression-based localization approach. Modeling was performed by a convolutional and recurrent neural network (CRNN) with two output branches, one performing SED and the other localization. In the localization branch, one regressor per class returned continuous azimuth and elevation angles. Binary thresholding was used in the detection branch to indicate the temporal activity of each class and that output was used to gate the respective direction-of-arrival (DoA) output, joining them together during inference. The proposed system, named SELDnet, was extensively compared against other architectures, for a variety of simulated and real data and for different array configurations. Note that both DNN-based proposals were using simple generic input features, such as multichannel power spectrograms in~\cite{hirvonen2015classification} or magnitude and phase spectrograms in~\cite{adavanne2018sound}.

Due to its relevance in the aforementioned applications, the SELD task was introduced for the first time in the DCASE 2019 Challenge and received a remarkable number of submissions for a novel topic. A new dataset of spatialized sound events was generated for the task 
\cite{adavanne2019multi} and a SELDnet implementation was provided by the authors as a baseline for the challenge participants\footnote{\url{https://github.com/sharathadavanne/seld-dcase2019}}. 
Beyond the works associated with the challenge \cite{Kapka2019, Cao2019, Xue2019_report, He2019_report, Pratik2019, Nguyen2019_report, MazzonYasuda2019_report,Chang2019_report,Ranjan2019,Park2019a,Leung2019_report,Grondin2019,ZhaoLu2019_report,Rough2019_report,Tan2019_report,CordourierMaruri2019,Krause2019_report,PerezLopez2019,Chytas2019,Anemueller2019_report,Kong2019_report,Lin2019_report}, multiple works have followed aiming to address the SELD task in a new way or improve on the limitations of the challenge submissions \cite{trowitzsch2019joining, pi2020u, wang2019localization, nguyen2020sequence}.

This paper serves three major aims. Firstly, it presents an overview of the first SELD-related challenge. Secondly, it presents common considerations of SELD systems and discusses how these were addressed by the participants, highlighting novel solutions and common elements of the challenge submissions. Thirdly, the performance of the systems is analyzed by addressing the issue of evaluating joint detection and localization. Following the ranking of the systems in the challenge, we calculate confidence intervals for the challenge evaluation metrics and analyze submissions with respect to their performance in detection and localization separately. Additionally, we reevaluate the systems using novel metrics proposed for joint evaluation of localization and detection \cite{Mesaros_2019_WASPAA} and investigate correlations between the different metrics and the ranking of the systems.

The paper is organized as follows: Section \ref{sec:task_descr} presents the task description, dataset, baseline system, and evaluation, as defined in the challenge. Section \ref{sec:seld_metric} introduces and formulates the joint metrics for evaluation of localization and detection. Section \ref{sec:analysis} presents the analysis of submitted systems, including the challenge results and detailed systems characteristics. In Section \ref{sec:reeval} we reevaluate the submissions with the new joint metrics, and analyze the results with a rank correlation analysis of the different metrics. Finally, Section \ref{sec:concl} presents the concluding remarks on the challenge task organization.  


\section{Sound event detection and localization in DCASE 2019 Challenge}
\label{sec:task_descr}
\begin{figure}
    \centering
    \includegraphics[width=0.9\columnwidth]{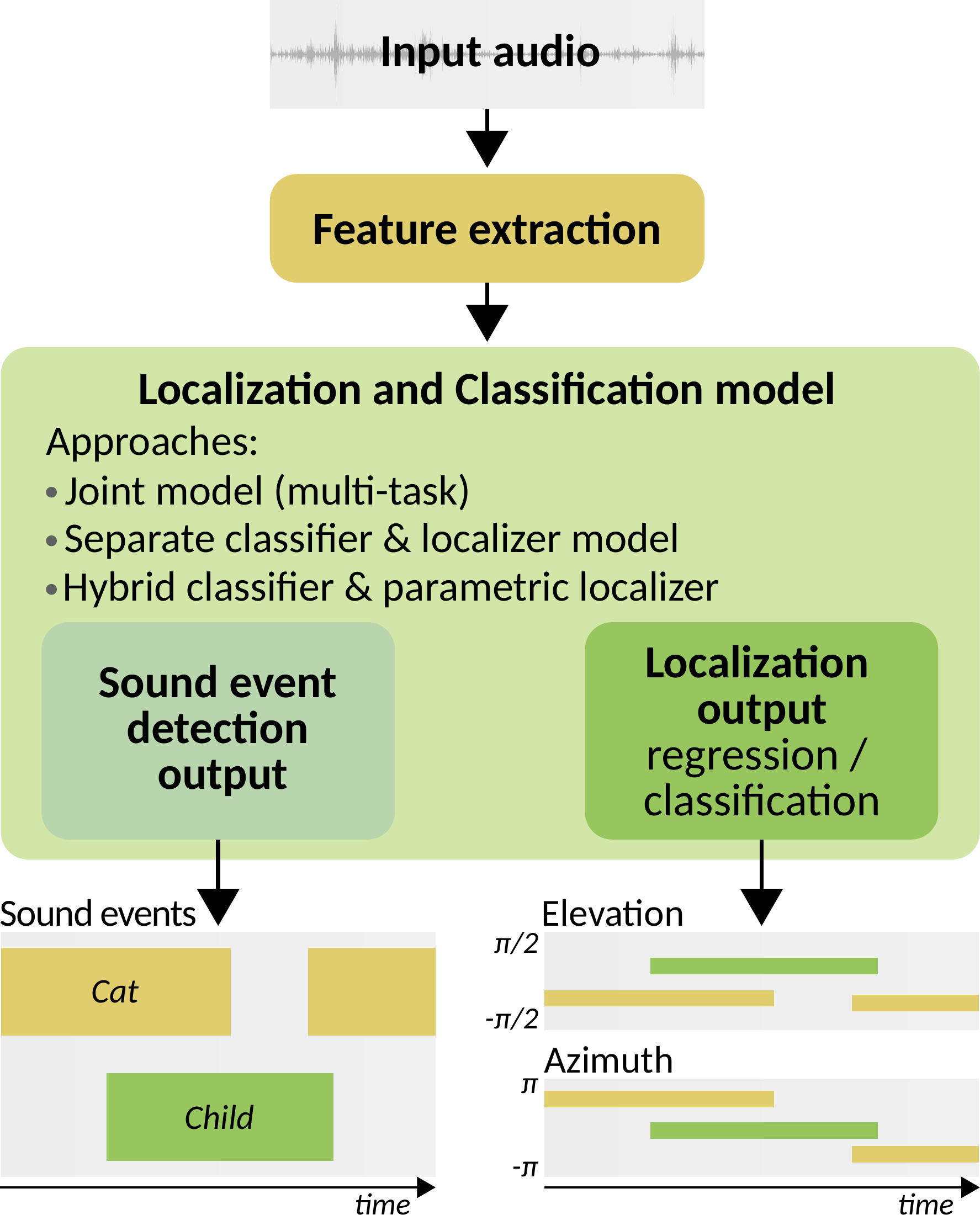}
    \caption{General SELD system approaches common in the challenge.}
    \label{fig:seld_diagram1}
    \vspace{-15pt}
\end{figure}

The goal of the SELD task, given a multichannel recording, can be summarized as identifying individual sound events from a set of given classes, their temporal onset and offset times in the recording, and their spatial trajectories while they are active. In the 2019 challenge, the spatial parameter was the DoA in azimuth and elevation, and only static scenes were considered, meaning that each individual sound event instance in the provided recordings was spatially stationary with a fixed location during its entire duration. Some common approaches to SELD systems found in the challenge are depicted in Fig.~\ref{fig:seld_diagram1}, including a single DNN modeling jointly the class and location of events, separate DNNs for classification and localization, or systems combining DNN-based classification with parametric localization.

\subsection{Dataset}

Creating a dataset for a SELD task presents some challenges, reflecting the high complexity of the problem. Ideally, a large range of sound events representative of each sound class should be reproduced at different times and temporal overlaps, at an enormous range of different positions in azimuth, elevation, and possibly distance from the microphones, covering the localization domain of interest. Furthermore, if the system is to be robust to varying acoustic conditions and different spaces, all the previous dimensions should be varied across different rooms. Staging real recordings with this degree of variability is not practical. Acoustic simulations of spatial room impulse responses (RIRs) for various room shapes and positions, and then subsequent convolution of the sound event samples with them is a viable alternative, explored for example in \cite{adavanne2018sound}. However, such simulators, with simplifications on room geometry and acoustic scattering behavior, can deviate significantly from real spatial RIRs. Additionally, the non-directional ambient noise characteristic of the function of each space is present in reality, adding another component the SELD system should be robust to.

For DCASE2019, we opted for a hybrid recording-simulation strategy that allowed us to control the detection, localization, and acoustical variability we needed. Real-life impulse responses were recorded at 5 indoor locations in the Hervanta campus of Tampere University, at 504 unique combinations of azimuth-elevation-distance around the recording position. The measurements were covering a domain of 360$^{\circ}$ in azimuth, -40$^{\circ}\sim$ 40$^{\circ}$ in elevation, and 1$\sim$2m in distance. Additionally, realistic ambient noise was recorded on-site with the recording setup unchanged.

Each spatial sound recording was synthesized as a one-minute multichannel mixture of spatialized sound events convolved with RIRs from the same space, with randomized onsets and source positions, and with up to two simultaneous events allowed. The RIRs were convolved with the isolated sound events  dataset\footnote{\url{https://archive.org/details/dcase2016_task2_train_dev}} provided with DCASE 2016 Task 2 Sound event detection in synthetic audio\footnote{\url{http://dcase.community/challenge2016/task-sound-event-detection-in-synthetic-audio}}, containing 20 event samples for each of the 11 event classes. Finally, the recorded natural ambient noise from the same space was added to the synthesized mixture, at a 30 dB signal to noise ratio relative to the average power of the sound-event mixture at the array channels. Each mixture was provided in two different 4-channel recording formats, extracted from the same 32-channel recording equipment. 
The first was a tetrahedral microphone array of capsules mounted on a hard spherical body, while the second was the first-order Ambisonics (FOA) spatial audio format. The two recording formats offer different possibilities in exploiting the spatial information captured between the channels. A development set was available during the challenge\footnote{\url{https://zenodo.org/record/2580091}}, and for the evaluation set only the audio without labels was released\footnote{\url{https://zenodo.org/record/3066124}}. 
The development and evaluation sets consist of 400 and 100 one-minute recordings, respectively. Half of the material has no overlapping events, while the other half has two overlapping events active for most of its duration. Note that two simultaneous events of the same class can occur in the overlapping case.
A detailed description of the generation of the dataset is given in \cite{adavanne2019multi}.

\subsection{Baseline system}

\begin{figure}
    \centering
    \includegraphics[width=1.0\columnwidth]{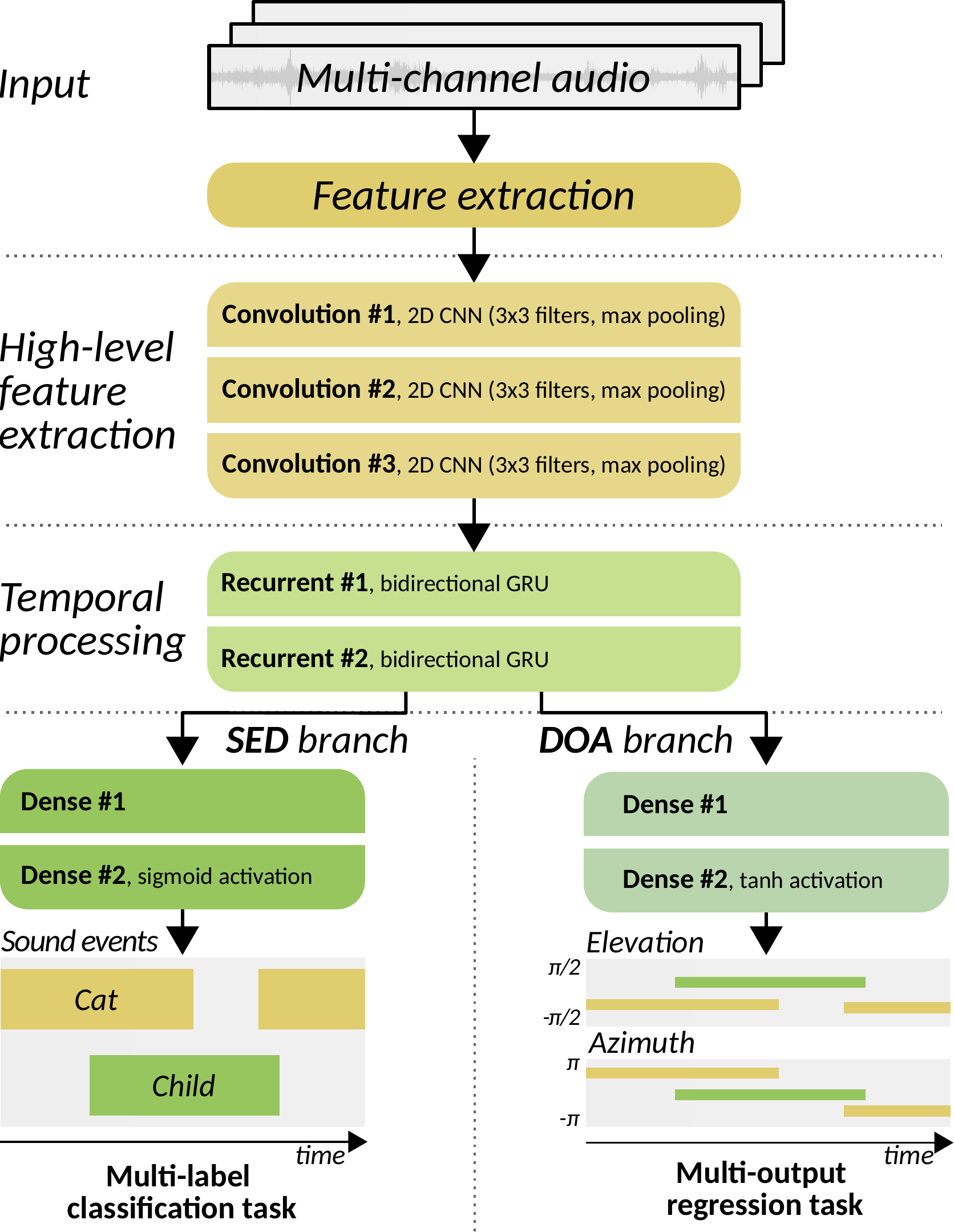}
    \caption{Detailed SELDnet network architecture of the baseline.}
    \label{fig:seld_diagram2}
    \vspace{-15pt}
\end{figure}

The SELDnet architecture of \cite{adavanne2018sound} was provided as the baseline architecture of the challenge. The rationale behind this choice was its conceptual and implementation simplicity, and its generality with respect to input features. Furthermore, even though SELDnet was very recent and had the best results between the tested methods in its publication, it still left a significant margin for improvements with realistic data, both at localization and detection accuracy. The architecture of the system is depicted in Fig.~\ref{fig:seld_diagram2}. It consists of three convolutional layers modeling spatial interchannel and sound event intrachannel time-frequency representations, followed by two bi-directional recurrent layers with gated recurrent units (GRU) capturing longer temporal dependencies in the data. The following two output branches of fully connected layers correspond to the individual tasks of SED and DoA estimation. The SED output is optimized with a cross-entropy loss, while the DoA output is optimized using the mean squared error of angular distances between reference and predicted DoAs. Contrary to the original SELDnet in \cite{adavanne2018sound} which was outputting Cartesian vector DoAs, the implementation for the challenge is returning directly azimuth and elevation angles. The network takes as input multichannel magnitude and phase spectrograms, stacked along the channel dimension. Reference SED outputs are expressed with one-hot encoding and reference DoAs with azimuth and elevation angles in radians. The network is trained using the Adam optimizer with a weighted combination of the two output losses, with more weight given to the localization loss. More details on the SELDnet challenge implementation can be found in \cite{adavanne2019multi}.

\subsection{Evaluation and ranking}
\label{ssec:eval_and_rank}
In this first implementation of the challenge the submitted systems were evaluated with respect to their detection and localization performance individually. For SED, the detection metrics were the $F1$-score and error rate ($ER$) computed in non-overlapping one-second segments \cite{Mesaros2016_MDPI}.
For DoA estimation, two additional frame-wise metrics were used. The first is a conventional \emph{directional error} ($DE$) expressing the angular distance between reference and predicted DoAs. Since multiple simultaneous estimates are possible, references and predictions need to be associated before errors can be computed. The Hungarian algorithm \cite{kuhn1955hungarian} was used for that purpose, and the final $DE$ was computed as the minimum cost association, divided with the number of associated DoAs. Since $DE$ does not reflect on how successfully a system detects localizable events, a second recall-type metric was introduced, termed \emph{frame recall} ($FR$). Due to a more general introduction and reformulation of the metrics, $DE$ is renamed in this work as \emph{localization error} ($LE$), while $FR$ is renamed as \emph{event count recall} ($ECR$). 

For a detailed picture of the overall performance, the submissions were ranked individually for each of the four $(F1, ER, LE, ECR)$ metrics. 
Hence, the $j$th submission had ranks $I_{F1}(j), I_{ECR}(j)$ based on its position after sorting the $F1, ECR$ results in descending order, and ranks $I_{ER}(j), I_{LE}(j)$ after sorting the $ER, LE$ results in ascending order.
A total ranking $I_\mathrm{tot}(j)$ aiming to indicate systems achieving good performance in all metrics or exceptional performance in most of them, was obtained by summing the individual ranks $I_{F1}(j)+I_{ECR}(j)+I_{ER}(j)+I_{LE}(j)$ and sorting the results in increasing order.

\section{Joint measurement of localization and detection performance}
\label{sec:seld_metric}

\begin{figure}
\centering
\includegraphics[width=1.0\columnwidth]{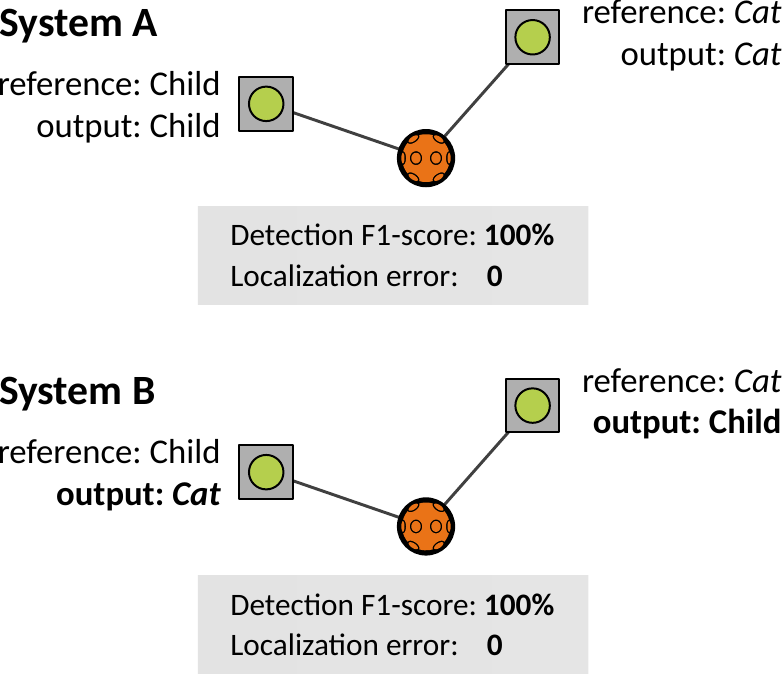}
\caption{Example reference and predicted sound events and locations. Circles denote reference sounds, rectangles system output. Two systems evaluated separately for detection and localization performance. Based on the measured performance, they both have perfect score.} 
\label{fig:motiv}
\end{figure}

Sound localization and sound event detection are traditionally two different areas of research, but the recent research addresses joint modeling and prediction of the two, motivating a joint evaluation. An example case to illustrate the main drawback of employing separate evaluations for detection and localization (similar to Subsection~\ref{ssec:eval_and_rank}) is visualized in Fig.~\ref{fig:motiv}. Both the participating systems have detected the two sound events correctly, however, their spatial positions are swapped. 
Using a standalone detection metric will evaluate if the system has correctly predicted the presence of the sound events (without regard to their position), and similarly, a standalone localization metric will evaluate the spatial errors between the closest sound pairs (ignoring the underlying sound classes). Hence, those metrics individually give the exact same score for both systems A and B, even though it is obvious that system B is inferior to A.

\subsection{Metrics formulation}

Since a spatial event is not distinguished only by its class, but also by its location, measurement ideally happens at the event level. Let us consider a SELD system that at a given temporal step predicts a set of $M$ events $P = \{p_1, ...,p_i, ..., p_{M}\}$, where each event prediction is associated with a class label index $\bp_i$ and a positional vector $\xp_i$, such that $p_i = \{\bp_i, \xp_i\}$. At the same time, $N$ reference events exist as $R = \{r_1, ...,r_j, ..., r_{N}\}$, with each reference event being of class index $b_j$ at position $\xr_j$, denoted as $r_j = \{b_j,\xr_j\}$. We assume a total of $C$ possible class labels that are ordered, such that $b\in[1,..,C]$. Note that contrary to traditional SED, where predictions and references are class based, it is possible that more than one of the events in $P$ or $R$ are of the same class.

We begin by considering localization-only metrics, neglecting classification. 
Every combination of prediction $\xp_i$ and reference $\xr_j$ is associated spatially with an appropriate distance metric $d(\xp_i,\xr_j)$. Two most common examples are
\begin{align}
    d(\xp_i,\xr_j) &=
        \arccos\left(\frac{\xp_i \cdot \xr_j}{||\xp_i|| ||\xr_j||}\right) & \mathrm{angular\; distance} \\
    d(\xp_i,\xr_j) &= 
        ||\xp_i - \xr_j|| &\mathrm{Cartesian\; distance}
\end{align}
In this work evaluation is based on angular distances since only directions of events are measured, instead of absolute positions.
All such distances can be expressed with an $M\times N$ distance matrix $\D$, where each element is given by $[\D]_{ij} = d(\xp_i,\xr_j)$. Before measuring a mean $LE$ across events, references and predictions should be associated using, for example, a minimum cost assignment algorithm such as the Hungarian algorithm, $\A = \mathcal{H}(\D)$. 
The association should ensure that if predictions are more than the references $M>N$, only $N$ predictions are associated with $N$ references, while if predictions are less than the references $M<N$, only $M$ references are associated with $M$ predictions. The non-associated $M-N$ predictions in the first case or the non-associated $N-M$ references in the second case, would constitute an equal number of false alarms (false positives $FP$) or misses (false negatives $FN$) respectively, in a detection-like localization metric. Consequently, if there are no predictions, all references are counted as $N$ misses, and similarly if there are no references, all predictions are counted as $M$ false alarms, while in such cases the $LE$ is undefined.
Based on the above, the $M\times N$ binary association matrix $\A$ can have maximum one unity entry at each column and row, meaning that only $K = \min(M,N) = ||\A||_1$ predictions and references are associated and contribute to the $LE$
\begin{equation}
    LE = \frac{1}{K}\sum_{i,j} a_{ij} d_{ij} = \frac{|| \A \odot \D||_1}{||\A||_1},
    \label{eq:le}
\end{equation}
where $d_{ij} = [\D]_{ij}$, $a_{ij} = [\A]_{ij}$, $||\cdot||_1$ is the $L_{1,1}$ entrywise matrix norm, and $\odot$ the entrywise matrix product.

The above localization precision gives a partial performance picture because it does not take into account misses or false alarms of localized sounds. To that purpose, we introduce a simple metric termed \emph{localization recall} ($LR$), expressed as
\begin{equation}
    LR = \frac{\sum_l \min(M^{(l)},N^{(l)})}{\sum_l N^{(l)}} = \frac{\sum_l ||\A^{(l)}||_1}{\sum_l N^{(l)}},
    \label{eq:lr}
\end{equation}
where the summation $\sum_l$ happens across $l=1,...,L$ temporal frame outputs or some other preferred averaged segmental representation, and $M^{(l)}, N^{(l)}$ are the number of predictions and references at the $l$th frame or segment.
Finally, a related but more concentrated metric of interest may be the ratio of frames or segments for which the system detects the correct number of references $M=N$. We name this metric \emph{event count recall} ($ECR$). $ECR$ corresponds to 
\begin{equation}
    ECR = \frac{\sum_l \mathbb{1}\left( M^{(l)} = N^{(l)} \right)} { L },
    \label{eq:ecr}
\end{equation}
and $\mathbb{1}(\cdot)$ is the indicator function, returning 1 if its argument is true, and 0 otherwise.
Note that $ECR$ was termed \emph{frame recall} in the challenge evaluation, and in \cite{adavanne2018direction, adavanne2018sound}, but we opted here for a more descriptive name of its counting objective.

Often, a localization method needs to be evaluated only under a certain level of spatial precision, usually expressed through an application-dependent threshold $\Theta$. Such a threshold on the above metrics can be applied by constructing an $M\times N$ binary matrix $\T$ with unity entries only on the reference-prediction pairs that are closer than the threshold, $[\T]_{ij} = \mathbb{1}([\D]_{ij}\leq \Theta)$. The number of associated predictions that pass the threshold are then given by $K_{\leq\Theta} = ||\T\odot\A||_1$. The thresholded metrics are
\begin{align}
    LE_{\leq\Theta} &= \frac{1}{K_{\leq\Theta}}\sum_{i,j} t_{ij} a_{ij} d_{ij} = \frac{||\T\odot\A\odot\D||_1}{||\T\odot\A||_1} \label{eq:le_theta}\\
    LR_{\leq\Theta} &= \frac{\sum_l K_{\leq\Theta}^{(l)}}{\sum_l N^{(l)}} = \frac{\sum_l ||\T^{(l)}\odot\A^{(l)}||_1}{\sum_l N^{(l)}}  \label{eq:lr_theta}\\
    ECR_{\leq\Theta} &= \frac{\sum_l \mathbb{1}\left( K_{\leq\Theta}^{(l)} = N^{(l)} \right)} { L } \label{eq:ecr_theta},
\end{align}
with $t_{ij} = [\T]_{ij}$. Note that the thresholded $LE_{\leq\Theta}$ in Eq.~\eqref{eq:le_theta} is undefined when there are no associations passing the threshold $K_{\leq\Theta}=0$.

Considering the fact that events have a class label in SELD, it is more informative to measure localization performance only between events that are correctly classified (class-aware localization). Similarly, we may want to impose a spatial constraint on correct classifications, such that events classified correctly, but very far from their spatial reference are considered invalid (location-aware detection). For both modes, we:
\begin{enumerate}
    \item Find subsets $P_c = \{p_i | \bp_i=c\}$ of predictions and $R_c = \{r_j | b_j=c\}$ of reference events classified on class $c\in[1,...,C]$. The resulting class-specific number of predictions is $M_c$ and of references $N_c$.
    \item Compute a class-dependent $M_c\times N_c$ distance matrix $\D_c$ between predictions $P_c$ and references $R_c$, and compute the respective association matrix $\A_c = \mathcal{H}(\D_c)$. 
    \item Determine a suitable application-specific spatial threshold $\Theta$, for location-aware detection. Construct the thresholding binary matrix $\T_c$ from $\D_c$, and determine the number of associated predictions $K_c = ||\A_c||_1 = \min(M_c,N_c)$, and the number of associated predictions which pass the threshold $K_{c,\leq\Theta} = ||\T_c\odot\A_c||_1$. 
    \item After association, count true positives $TP$, false negatives $FN$, and false positives $FP$ as follows:
\end{enumerate}    
    \begin{align}
        TP_{c,\leq\Theta} &= K_{c,\leq\Theta} \\
        FP_{c,\leq\Theta} &= \max(0,M_c-N_c) + \min(M_c, N_c)-K_{c,\leq\Theta} \\
        FN_{c} &= \max(0,N_c-M_c).
    \end{align}
 
\begin{figure}
 \centering
 \includegraphics[width=\columnwidth]{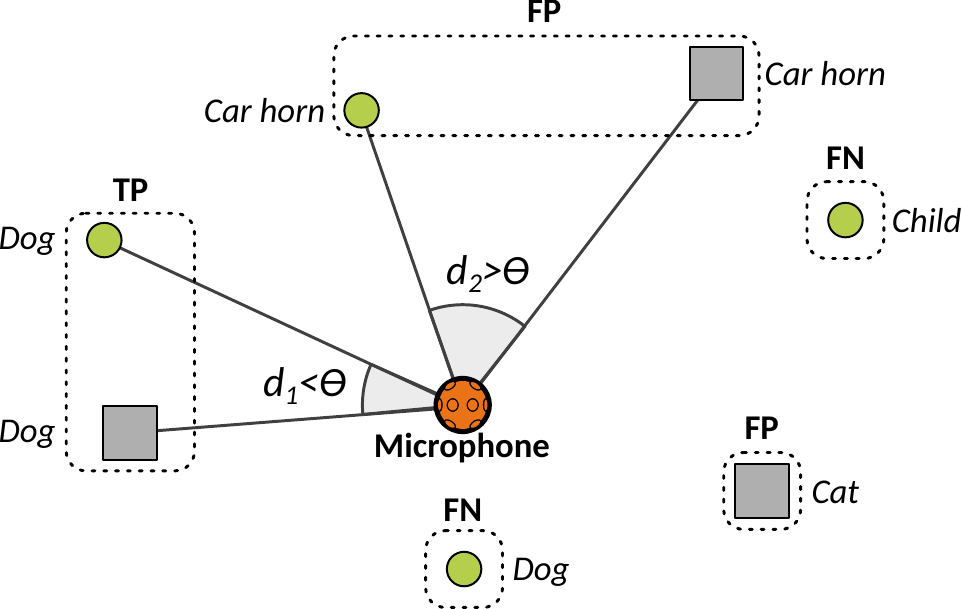}
 \caption{Example reference and predicted sound events and locations. Circles denote reference sounds, rectangles system output. The dashed rectangles indicate associated predictions with references, and $d_1, d_2$ are the respective angular distances expressing localization error. The $TP,FP,FN$ indicate how the respective predictions and references contribute to the true positive, false positive, and false negative count.}
 \label{fig:motiv-2}
 \end{figure}

A simple example is illustrated in Fig.~\ref{fig:motiv-2}, where the reference annotation contains four sound events: \textit{dog}, \textit{dog}, \textit{car horn}, and \textit{child}, while the system output contains three: \textit{dog}, \textit{car horn}, and \textit{cat}, at their respective positions. The joint evaluation will compare for correctness of both the labels and the locations, therefore it will characterize the localization error in the \textit{dog}-\textit{dog} pair and the \textit{car horn}-\textit{car horn} pair, and consider the other events as errors (false positives and false negatives). 
Note that with the above setup false negatives do not depend on the threshold, while false positives include both the extraneous predictions and associated predictions that did not pass the threshold (the \textit{car horn} example in Fig.~\ref{fig:motiv-2}). Based on the above, we are able to measure location-aware detection metrics such as precision, recall, F1-score, or error rates.
    
Regarding class-aware localization, we compute the localization error ($LE_c$) and localization recall ($LR_c$) of Eq.~\eqref{eq:le}--\eqref{eq:lr} only between predictions and references of class $c$
    \begin{align}
        LE_{c} &= \frac{||\A_c\odot\D_c||_1}{||\A_c||_1} \label{eq:le_c}\\
        LR_{c} &= \frac{\sum_l ||\A_c^{(l)}||_1}{\sum_l N_c^{(l)}} \label{eq:lr_c}.
    \end{align}
    The overall \emph{class-dependent} $LE_{CD}, LR_{CD}$, are computed as the class means of Eq.~\eqref{eq:le_c}--\eqref{eq:lr_c}
    \begin{align}
        LE_{CD} &= \frac{1}{C\cdot L} \sum_c \sum_l LE_{c}^{(l)} \label{eq:le_cd}\\
        LR_{CD} &= \frac{1}{C} \sum_c LR_{c} \label{eq:lr_cd}.
    \end{align}    
In some applications it may be of interest to have both class-dependent, and thresholded localization metrics, similar to Eq.~\eqref{eq:le_theta}--\eqref{eq:ecr_theta}. In the joint measurement results of this study we use the non-thresholded versions of Eq.~\eqref{eq:le_c}--\eqref{eq:lr_c}. It is also worth noting that different thresholds per class $\Theta_c$ may be accommodated in the above framework, to reflect different spatial tolerances for certain classes depending on the application. 
In our evaluation we opted for non-thresholded localization metrics since we deemed it more beneficial to have a localization measurement of all detected estimates in a class, providing complementary information to the spatially-thresholded detection metrics.

It is worth noting here the relation between the proposed metrics and dedicated tracking metrics such as the OSPA \cite{schuhmacher2008consistent} or the CLEAR MOT \cite{bernardin2008evaluating} metrics, which evaluate the performance of systems in identifying distinct contiguous spatial trajectories from instantaneous spatial estimates. A form of tracking occurs in SELD systems through location-aware classification; the positional estimates of a source emitting a signal of a certain type are joined to a consistent spatiotemporal trajectory when that class is detected active. However, tracking metrics evaluate trajectory consistency without having to resort to classes and they penalize identity switches, something that the proposed metrics do not do e.g. in the case of two simultaneous events of the same class.

\subsection{Segment-based measurement}

Segment-based detection metrics generalize the frame-based binary activity of sound events to its corresponding activity at segment-level and are common in SED. In~\cite{Mesaros2016_MDPI}, this generalization is done by considering an event to be active at segment-level if it is active in at least one frame within the segment. 
A similar generalization of the localization metrics to a different time-scale can be formulated through a spherical mean DoA vector or Cartesian mean positional vector $\hat{\mathbf{x}}$ of all predictions $\xp^{(l)}$ of the corresponding event within the segment, before localization errors are measured. Alternatively, the average localization error within a segment can be computed based on the frame-based pairs of reference and predicted events. Both approaches are introduced and compared in \cite{Mesaros_2019_WASPAA} with comparable results. 
Herein, we present results evaluated in one-second segments, apart from the reevaluation in Sec.~\ref{sec:reeval} where additional frame-level localization results are included in the analysis.

\section{Challenge results}
\label{sec:analysis}

Even though the SELD task was introduced in DCASE2019 for the first time, it attracted a lot of interest and received the second highest number of submissions among other tasks. In total 58 systems were submitted, from a total of 22 teams consisting of in total 65 members. The participants were affiliated with 16 universities and 8 companies.

\begin{table*}
\centering
\caption{Challenge results of submitted systems. The rank is based on the cumulative rank based on the four calculated metrics. Best system per team according to the official challenge ranking. Best score indicated for the separate metrics.}
\begin{tabular}{r l c c c c}
\toprule
\label{tab:official-results}
Rank & System & ER & F1 & LE & ECR \\
\midrule
1 & Kapka\_SRPOL\_2~\cite{Kapka2019} 
& 0.08  $\pm$ 0.01   & 94.7  $\pm$ 0.8 & 3.7  $\pm$ 0.6    & \textbf{96.8  $\pm$ 0.6}    \\

2 & Cao\_Surrey\_4~\cite{Cao2019}  
& 0.08 $\pm$ 0.01   & 95.5 $\pm$ 0.4    & 5.5  $\pm$ 0.7    & 92.2  $\pm$ 1.0    \\

3 & Xue\_JDAI\_1~\cite{Xue2019_report} 
& \textbf{0.06 $\pm$ 0.01}  & 96.3 $\pm$ 0.5    & 9.7  $\pm$ 1.3 & 92.3  $\pm$ 1.3    \\

4 & He\_THU\_2~\cite{He2019_report}   
& \textbf{0.06 $\pm$ 0.01}   & \textbf{96.7 $\pm$ 0.4}    & 22.4 $\pm$ 1.7    & 94.1  $\pm$ 1.0    \\

5 & Jee\_NTU\_1~\cite{Pratik2019}  
& 0.12 $\pm$ 0.01   & 93.7 $\pm$ 0.5    & 4.2  $\pm$ 0.5    & 91.8  $\pm$ 1.0    \\

6 & Nguyen\_NTU\_3~\cite{Nguyen2019_report}  
& 0.11 $\pm$ 0.01   & 93.4 $\pm$ 0.7    & 5.4  $\pm$ 0.4    & 88.8  $\pm$ 1.6    \\

7 & MazzonYasuda\_NTT\_3~\cite{MazzonYasuda2019_report} 
& 0.10 $\pm$ 0.01   & 94.2 $\pm$ 0.5    & 6.4  $\pm$ 0.9    & 88.8  $\pm$ 1.3    \\

8 & Chang\_HYU\_3~\cite{Chang2019_report}  
& 0.14 $\pm$ 0.01   & 91.9 $\pm$ 0.5 & \textbf{2.7  $\pm$ 0.3}    & 90.8  $\pm$ 1.3    \\

9 & Ranjan\_NTU\_3~\cite{Ranjan2019}  
& 0.16 $\pm$ 0.01   & 90.9 $\pm$ 0.8    & 5.7  $\pm$ 0.5    & 91.8  $\pm$ 1.0    \\

10 & Park\_ETRI\_1~\cite{Park2019a} 
& 0.15 $\pm$ 0.01   & 91.9 $\pm$ 0.6    & 5.1  $\pm$ 0.7    & 87.4  $\pm$ 1.3    \\

11 & Leung\_DBS\_2~\cite{Leung2019_report} 
& 0.12 $\pm$ 0.01   & 93.3 $\pm$ 0.6    & 25.9 $\pm$ 1.3    & 91.1  $\pm$ 1.3    \\

12 & Grondin\_MIT\_1~\cite{Grondin2019} 
& 0.14 $\pm$ 0.01   & 92.2 $\pm$ 0.7    & 7.4  $\pm$ 0.6    & 87.5  $\pm$ 1.7    \\

13 & ZhaoLu\_UESTC\_1~\cite{ZhaoLu2019_report}    
& 0.18 $\pm$ 0.01   & 89.3 $\pm$ 0.8    & 6.8  $\pm$ 0.9    & 84.3  $\pm$ 1.4    \\

14 & Rough\_EMED\_2~\cite{Rough2019_report}  
& 0.18 $\pm$ 0.01   & 89.7 $\pm$ 0.7    & 9.4  $\pm$ 0.9    & 85.5  $\pm$ 1.5    \\

15 & Tan\_NTU\_1~\cite{Tan2019_report}  
& 0.17 $\pm$ 0.02   & 89.8 $\pm$ 0.9    & 15.4 $\pm$ 1.4    & 84.4  $\pm$ 2.1    \\

16 & Cordourier\_IL\_2~\cite{CordourierMaruri2019}   
& 0.22 $\pm$ 0.01   & 86.5 $\pm$ 0.8    & 20.8 $\pm$ 1.2    & 85.7  $\pm$ 1.5    \\

17 & Krause\_AGH\_4~\cite{Krause2019_report}  
& 0.22 $\pm$ 0.02   & 87.4 $\pm$ 0.9    & 31.0 $\pm$ 1.0    & 87.0  $\pm$ 1.8    \\ 

\rowcolor{Gray}
18 & Adavanne\_TAU\_FOA~\cite{adavanne2019multi}  
& 0.28 $\pm$ 0.02   & 85.4 $\pm$ 0.9    & 24.6 $\pm$ 1.1    & 85.7  $\pm$ 1.9    \\

19 & Perezlopez\_UPF\_1~\cite{PerezLopez2019}  
&0.29 $\pm$ 0.03   & 82.1 $\pm$ 1.5    & 9.3  $\pm$ 0.4    & 75.8  $\pm$ 2.5    \\

20 & Chytas\_UTH\_1~\cite{Chytas2019}  
& 0.29 $\pm$ 0.01   & 82.4 $\pm$ 0.8    & 18.6 $\pm$ 1.3    & 75.6  $\pm$ 2.4    \\

21 & Anemueller\_UOL\_3~\cite{Anemueller2019_report}  
& 0.28 $\pm$ 0.02   & 83.8 $\pm$ 1.2    & 29.2 $\pm$ 1.1    & 84.1  $\pm$ 2.3    \\

22 & Kong\_SURREY\_1~\cite{Kong2019_report}
& 0.29 $\pm$ 0.01   & 83.4 $\pm$ 0.9    & 37.6 $\pm$ 1.7    & 81.3  $\pm$ 1.9    \\

23 & Lin\_YYZN\_1~\cite{Lin2019_report} 
& 1.03 $\pm$ 0.01   & 2.6  $\pm$ 0.7    & 21.9 $\pm$ 8.2    & 31.6  $\pm$ 2.5   \\
\bottomrule
\end{tabular}
\end{table*}

\begin{table*}
\centering
\caption{Summary of submitted systems. The rank is based on the cumulative rank based on the four calculated metrics. Best system per team according to the official challenge ranking.}
\begin{tabular}{r l l l l c}
\toprule
\label{tab:official-results-sys-char}
 & System & Audio & Features & Classifier & Multi-task\\
\midrule
1 & Kapka\_SRPOL\_2~\cite{Kapka2019} 
& AMB  & Phase and magnitude spectra  & CRNN & $\times$
\\

2 & Cao\_Surrey\_4~\cite{Cao2019}  
& Both & Log-mel, GCC, and intensity vectors &  CRNN ensemble  & $\times$
\\

3 & Xue\_JDAI\_1~\cite{Xue2019_report} 
& MIC & Log-mel, Q-transform, multiple spectra &  CRNN ensemble, parametric DoA & \checkmark \\

4 & He\_THU\_2~\cite{He2019_report}   
& AMB & Log-mel, phase, and magnitude spectra &  CRNN  & $\times$
\\

5 & Jee\_NTU\_1~\cite{Pratik2019}  
& MIC & Log-mel spectra and GCC
 &   CRNN & $\times$
\\

6 & Nguyen\_NTU\_3~\cite{Nguyen2019_report}  
& AMB & Log-mel, phase, and magnitude spectra &  CRNN, parametric DoA  & $\times$
\\

7 & MazzonYasuda\_NTT\_3~\cite{MazzonYasuda2019_report} 
& Both & Log-mel spectra and GCC &   CRNN, ResNet ensemble  & $\times$
\\

8 & Chang\_HYU\_3~\cite{Chang2019_report}  
& MIC & Log-mel spectra, cochleagram, and GCC & CRNN, CNN  & $\times$
\\

9 & Ranjan\_NTU\_3~\cite{Ranjan2019}  
& MIC  &Log-mel and phase spectra  & ResNet RNN  & $\times$
\\

10 & Park\_ETRI\_1~\cite{Park2019a} 
& Both & Log-mel and intensity vectors & CRNN, TrellisNet & \checkmark
\\

11 & Leung\_DBS\_2~\cite{Leung2019_report}  
& AMB & Log-magnitude, phase, and cross spectra & CRNN ensemble  & \checkmark
\\

12 & Grondin\_MIT\_1~\cite{Grondin2019}   
& MIC & Phase and magnitude spectra, GCC and TDOA  & CRNN ensemble & $\times$
\\

13 & ZhaoLu\_UESTC\_1~\cite{ZhaoLu2019_report} 
& MIC & Log-mel spectra & CRNN  & \checkmark   
\\

14 & Rough\_EMED\_2~\cite{Rough2019_report}  & MIC  & Phase and magnitude spectra  & CRNN & $\times$
\\

15 & Tan\_NTU\_1~\cite{Tan2019_report}  
& MIC & Log-mel spectra and GCC  & ResNet RNN, parametric DoA  & $\times$
\\

16 & Cordourier\_IL\_2~\cite{CordourierMaruri2019}   
& MIC & Phase and magnitude spectra, and GCC  & CRNN ensemble & \checkmark
\\

17 & Krause\_AGH\_4~\cite{Krause2019_report}  
& AMB  &  Phase and magnitude spectra & CRNN ensemble & \checkmark
\\ 
\rowcolor{Gray}
18 & Adavanne\_TAU\_FOA~\cite{adavanne2019multi}  
& AMB &  Phase and magnitude spectra & CRNN & \checkmark
\\

19 & Perezlopez\_UPF\_1~\cite{PerezLopez2019}  
& AMB & Log-mel spectra  & CRNN, parametric DoA & $\times$
\\

20 & Chytas\_UTH\_1~\cite{Chytas2019}  
& MIC & Raw audio and power spectra & CNN ensemble & $\times$
\\

21 & Anemueller\_UOL\_3~\cite{Anemueller2019_report}  
& AMB & Group-delay and magnitude spectra & CRNN & \checkmark
\\

22 & Kong\_SURREY\_1~\cite{Kong2019_report} 
& AMB & Magnitude spectra  & CNN  & \checkmark
\\

23 & Lin\_YYZN\_1~\cite{Lin2019_report} 
& AMB & Phase and magnitude spectra & CRNN  & \checkmark
\\
\bottomrule
\end{tabular}
\end{table*}

\subsection{Overall challenge results}

The overall results of the challenge are presented in Table \ref{tab:official-results}. Only the best system of each team is presented and the systems are ordered by their official challenge rank as described in Section~\ref{ssec:eval_and_rank}. In addition to the results displayed on the challenge webpage, this table includes the 95\% confidence intervals for each separate metric, estimated using the jackknife procedure presented in \cite{Mesaros2019_TASLP}. The method is a resampling technique that estimates a parameter from a random sample of data for a population using partial estimates \cite{abdi2010jackknife}. Confidence intervals by jackknifing are coarse approximations, but applicable in cases where the underlying distribution of the parameter to be estimated is unknown. In our case the parameters are metrics that depend on individual combinations of active sounds at each time and the jackknife method allows estimating the confidence intervals without making any assumption on their distribution. The partial estimates for all metrics were calculated in a leave-one-out manner, excluding, in turns, one audio file from the evaluation set. 

Considering the best-performing system of each team, 17 out of the 22 submitted systems ranked higher than the baseline system using the official ranking method. In terms of the individual metrics, 17 systems had better $ER$ and $F1$-scores than the baseline, with the best $ER$ and $F1$-scores of 0.06~\cite{Xue2019_report, He2019_report} and 96.7\%~\cite{He2019_report} respectively. Similarly, 18 systems had better $LE$ and 14 systems had higher $ECR$, with the best $LE$ of 2.7$^\circ{}$~\cite{Chang2019_report} and $ECR$ of 96.8\%~\cite{Kapka2019}. 

\begin{figure*}[!t]
  \centering
  \subfloat[Detection results]{{\includegraphics[width=0.5\textwidth]{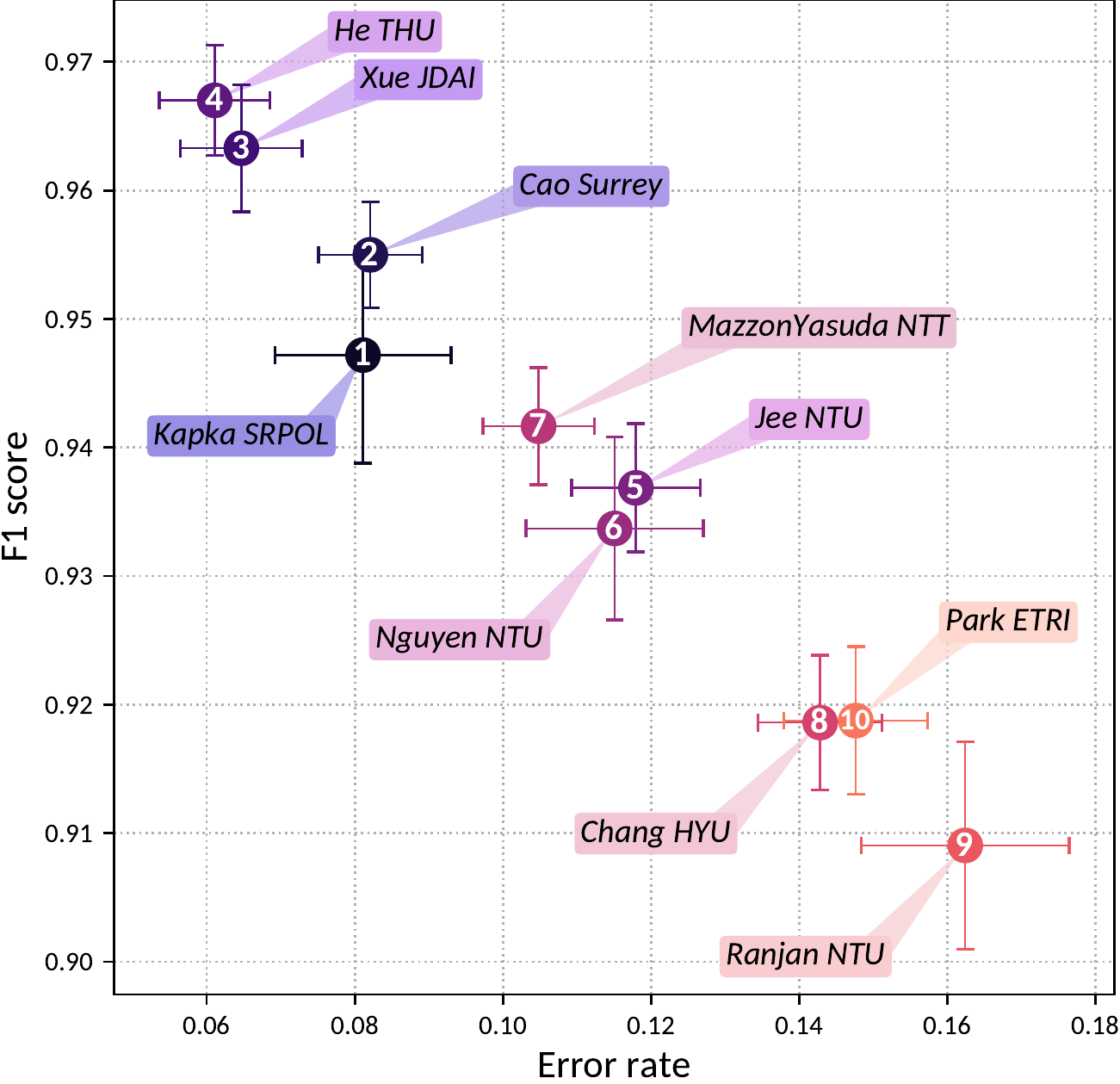} \label{fig:challenge-results-det}}}  
  \subfloat[Localization results]{{\includegraphics[width=0.5\textwidth]{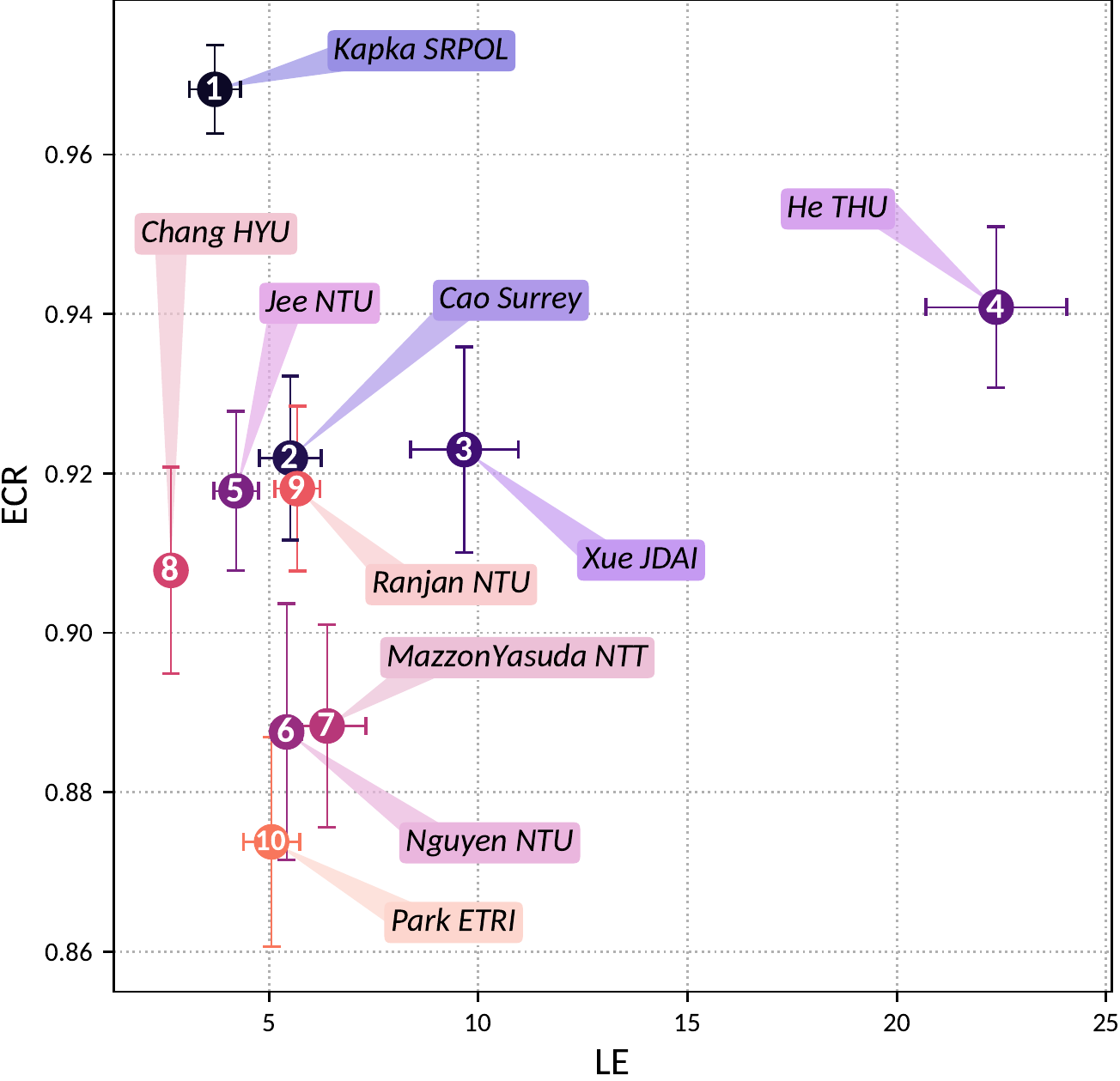} \label{fig:challenge-results-loc}}}
\caption{Separately calculated detection and localization performance of top 10 systems (best system per team). The official rank of the systems is indicated in the center of the marker for each scatter plot.}
\label{fig:challenge-results}
\end{figure*}

The top-10 systems of Table \ref{tab:official-results} are illustrated with respect to detection metrics in Fig. \ref{fig:challenge-results-det} and localization metrics in Fig. \ref{fig:challenge-results-loc}. The best system in both these plots is in their corresponding top left corner. We observe that the ranking order of the submitted systems is different for detection and localization metrics.
For instance, the best system according to detection metrics - \textit{He THU}~\cite{He2019_report} (Fig. \ref{fig:challenge-results-det} top-left corner), has a high $LE$ compared to the other top-10 systems and hence achieves an overall rank of four. Similarly, although \textit{Chang HYU}~\cite{Chang2019_report}
achieved the best $LE$ among the top-10 systems, its detection performance was among the poorest of top-10 systems and hence achieved a rank of eight. In general, $ER$ and $F1$-scores of event detection are correlated and hence all the submitted systems are observed along the diagonal. This diagonal behavior is not observed with the localization metrics as $LE$ and $ECR$ are only weakly correlated.

The system characteristics of all the submissions are summarized in Table~\ref{tab:official-results-sys-char}.
All systems had at least one deep learning component in their approach. Specifically, apart from~\cite{Chytas2019} and~\cite{Kong2019_report} that employed a CNN architecture with no recurrent layers the remaining 20 systems employed different versions of the baseline CRNN architecture as one of their components. 
Four of the submitted systems employed model-based parametric DoA estimation~\cite{Xue2019_report, Nguyen2019_report, PerezLopez2019, Tan2019_report} along with CRNN-based classification. The best purely parametric DoA approach~\cite{Nguyen2019_report} achieved the 6th position.
Among the DNN-based SELD methods, nine of them employed multi-task learning~\cite{caruana1997multitask} for joint SED and DoA estimation. The remaining systems, including the top ranked system~\cite{Kapka2019}, employed separate networks for SED and DoA estimation and performed engineered data-association of their respective outputs. Finally, there was no significant improvement in SELD performance with the choice of either of the two audio formats in the dataset. Among the top 10 ranked systems, four of them used the microphone array format, three used the Ambisonic format, and the rest used both formats as input.

\subsection{Analysis of individual systems}

A detailed analysis of some of the systems follows, along with a summary of the most prominent architectural, input feature, or training characteristics.

Kapka \& Lewandowski (\textit{Kapka SRPOL})\cite{Kapka2019} was the top performing system of the challenge, with very high performance in both localization and detection. There was minimal feature engineering and the pure magnitude and phase spectrograms of the FOA format were used as input. However, the approach was highly coupled to the task, by splitting it into four well defined subtasks and then dedicating one CRNN model to infer each one of them. The subtasks were: a) estimation of the number of sources, b) estimation of DoA for an active source, c) estimation of a second DoA in the case that two simultaneous events are detected, d) classification of events whose number equals the number of detected sources. 
Well-engineered post-processing of outputs, from source count to localization to event durations to classification, coupled the method to prior knowledge of the dataset and ensured consistent association and information flow between modules. It is worth noting that their architecture seems able to resolve two simultaneous instances of the same class at different directions. Since the architecture relied on prior knowledge, such as a maximum of two simultaneous sources and discrete DoAs at 10$^\circ$ intervals, it was not as general as most of the other approaches.

Cao et al. (\textit{Cao Surrey}) \cite{Cao2019}, had the second best performing system, following the first one closely. However, the authors kept the general SELDnet architecture and advanced it with a number of informed domain-specific choices. The most important ones seem to be improved input features and disassociation of the detection and localization losses. In particular, the losses were separated by duplicating the SELDnet and training each clone for SED and localization separately. Here, the ground truth SED activations were used as masks on the localization loss. Additionally, they used both FOA and MIC input and ensemble averaging. According to ablation studies in \cite{Cao2019}, the better input features and the two-stage training architecture have a drastic effect in performance.

The system of Xue et al. (\textit{Xue JDAI}) \cite{Xue2019_report} outperformed the first two in detection results, but had lower localization performance resulting in the third best average rank. Its success seems to be a combination of multiple spectral and spatial features and elaborate post-processing. DoA estimation from the CRNN model was also abandoned in favour of a traditional SRP estimation, refined by the former only in the case of simultaneous events. Additionally, separate CNN branches were used for SED and localization features, before being merged at the recursive layers.

The fourth best system of Zhang et al. (\textit{He THU}) \cite{He2019_report} follows the same architecture as \cite{Cao2019}. It had the best SED performance overall, but its localization accuracy was only marginally better than the baseline. The large difference compared to the second system may be due to the basic spectrogram feature for localization, instead of the more effective directional features used in \cite{Cao2019}. On the other hand, the higher detection performance may be attributed to the SpecAugment~\cite{park2019specaugment} data augmentation strategy used. The same architecture was also employed by the fifth best system of Jee et al. \cite{Pratik2019}, aiming to improve its performance. They introduced a number of incremental modifications to the SED features, CRNN layers, pooling, and activation functions, along with a mixup \cite{zhang2018mixup} data augmentation strategy, without, however, achieving better results at the challenge evaluation.

Nguyen et al. (\textit{Nguyen NTU}) \cite{Nguyen2019_report} took the concept of independent localization and detection to its extreme, performing them separately and then associating DoAs to overlapping detected events randomly. Good overall performance brought them to the sixth place. Note that their approach exploits the fact that detection and estimation performance are evaluated independently and correct associations between the two are not measured, as discussed in the next sections.

The next best system of Mazzon et al. (\textit{MazzonYasuda NTT}) \cite{MazzonYasuda2019_report} was also based on the architecture of the second best system \cite{Cao2019}, trying to improve on it with a Resnet network replacing CNNs, an elaborate ensemble strategy, and, most importantly, an original spatial data augmentation approach exploiting the rotation and reflection properties of the spherical harmonic bases encoding the sound field in Ambisonics \cite{Mazzon2019}. The authors limited the input features to only GCC-PHAT for both FOA and microphone array signals, potentially limiting their effectiveness for the FOA set which encodes DoA information by amplitude differences. 

Noh et al. (\textit{Chang HYU})~\cite{Chang2019_report} added an overall sound activity detection model on top of the SED one. Two additional independent CRNN models were trained to detect presence of one or two events respectively, using cochleagram features as input. Their binary outputs were used to select whether none, one, or two event classes with the highest probabilities of the dedicated CRNN SED model were outputted. The authors employed just a CNN network for DoA estimation, performed as a classification task on 324 classes, inferred from the grid of potential DoAs in the dataset. Interestingly, their model achieved the lowest localization error in the challenge. That may be attributed to their DoA classification matching the DoA discretized grid in the dataset, along with their spatial data augmentation technique, mixing recordings from non-overlapping events to generate additional overlapping segments for training. No information was provided on how or if DoAs were associated with events and from further analysis on the following sections we assume that such association was done randomly, as in \cite{Nguyen2019_report}. The same approach of independent SED and localization networks, a classification-based DoA estimation, and random association between the two was followed by the next best performing system of Ranjan et al. (\textit{Ranjan NTU})\cite{Ranjan2019}. Additionally, the authors replaced CNN layers with Resnets in the typical CRNN networks followed by most participants.

The tenth-best performing system of Park et al. (\textit{Park ETRI})\cite{Park2019a} attempted to combine the success of the two-stage training approach \cite{Cao2019} with the assumed consistency of joint-modeling. They performed two stages of weight transfer from separately trained SED and DoA estimation networks, into a new network with a SED and DoA branch trained with a combined detection and localization loss, as in the baseline SELDnet. Additionally they experimented with TrellisNet layers instead of RNNs and alternative activation functions.

We note some interesting investigations in the rest of the submitted systems. Grondin et al. (\textit{Grondin MIT}) \cite{Grondin2019} used one CRNN for each microphone pair in the array format, performing joint event detection and localization. The network was trained to output intermediate TDOA values, mapped afterwards to DoAs. Tan et al. (\textit{Tan NTU}) \cite{Tan2019_report} was one of the four systems that did not use machine learning for DoA estimation, computing time-domain cross-correlations between microphone pairs and their respective TDOA and converting it to a DoA by a least-squares geometric fit. Krause and Kowalczyk (\textit{Krause AGH}) \cite{Krause2019_report} explored various combinations of layers processing localization and SED features before fusion, as well as early branching for the two tasks.  \textit{Grondin MIT} \cite{Grondin2019} showed similar considerations on the fusion of  input features, since the approach of the baseline stacking phase and magnitude spectrograms into a single tensor could be suboptimal. Chytas and Potamianos (\textit{Chytas UTH})~\cite{Chytas2019} proposed to perform SELD directly from downsampled audio waveforms, with some additional help for SED using power spectrograms. Even though their CNN-only approach underperformed on SED, it showed that competitive localization can be achieved using DNNs directly on time-domain multichannel audio.

Finally, a special mention should go to the system by Perez et al. (\textit{PerezLopez UPF})\cite{PerezLopez2019} since, along with the best performing system of \cite{Kapka2019}, it was the only other system following a localize-before-detect paradigm. Their approach was based on model-based DoA estimation on the FOA format, determination of the number of sources based on the DoA estimates, determination of the event onset/offset, and beamforming towards the prominent DoAs. The beamformed signals, being essentially estimates of separated event signals, were fed to a CRNN classifier for SED. Contrary to the majority of submissions in the challenge, such an architecture is capable of detecting simultaneous instances of the same class localized at different directions.

\subsection{Discussion on submitted systems}

One obvious observation on the results is that the SELDnet baseline, as implemented for the challenge, had a suboptimal performance compared to the majority of the submissions. An initial weakness seems to be the input features. A number of submissions indicated that by switching to features with more concentrated information on each of the two tasks, detection (log-mel spectra) or localization (GCC-PHAT arrays, active intensity vectors), improved performance significantly. These three sets of features were the most popular overall in the top submissions, with only the third best system relying on multiple other types of multichannel spectra. It has to be noted though, that the top system \cite{Kapka2019} used the raw multichannel phase and magnitude spectrograms, indicating that it is possible to perform SELD succesfully with such lower level features, but with model architectures exploiting prior knowledge and coupled tightly to the task.

The most popular network architecture and training choices seem to be the ones introduced by Cao et al. \cite{Cao2019}. Essentially, their work disassociates the joint cost function combining SED losses and localization losses as realized in the baseline and trains individual models for each task. The SED and DoA estimates are then associated through a training strategy or assigned randomly between them \cite{Nguyen2019_report, Chang2019_report, Ranjan2019}. It has to be noted that such random association takes advantage of the fact that detection and localization were evaluated independently in the challenge and would not be a good strategy in practice. Ranjan et al. \cite{Ranjan2019} compared the two-stage architecture versus joint-modeling, with clearly improved results with the former. However, it is worth noting that two systems in the top ten places had a single network performing joint-modeling \cite{Xue2019_report, Park2019a}, one of them being third best \cite{Xue2019_report}.

The SELD paradigm proposed by the SELDnet baseline, where one DoA output is tied to each class, was followed by most submissions including multi-stage approaches \cite{Cao2019, Chang2019_report}. This paradigm 
is forcing a convenient detect-before-localize approach. However, it limits the output of the system to only one localized event per class, even in the presence of two same-class instances. Systems that were training an independent localization network as a DoA classification task, were not addressing that problem since association of DoAs to detected classes was ambiguous. The only two submissions that followed a localize-before-detect paradigm used localization information to determine the number and DoAs of events independently of their class \cite{Kapka2019, PerezLopez2019}. They were then passing that information to classifiers, turning the class-based outputs into event-based outputs and circumventing the same-class multi-instance problem of the detect-before-localize approach.

Certain architectural or training choices were specific to the localization task. Some of the submissions treated DoA estimation as a classification task \cite{Chang2019_report, Ranjan2019}, e.g. similar to other DNN-based localization works \cite{adavanne2018direction, perotin2019crnn, chakrabarty2019multi}, instead of the regression format of the baseline. Xue et al. \cite{Xue2019_report} trained both DoA output formats simultaneously. However, it has to be noted that the systems that relied only on DoA classification were taking advantage of the the small set of 324 fixed DoAs embedded in the dataset. A dataset with a much more dense spatial resolution of possible DoAs, a continuous range of DoAs, or moving sources, may have needed a much larger number of classes to be modeled effectively (e.g. 2522 discrete angles for a resolution of 5$^\circ$ in azimuth and elevation covering the sphere). Moreover, classification-based DoA estimation was found successful in two-stage systems, training independently a DoA network. Joint-modeling of SELD based completely on classification, as pioneered by Hirvonen \cite{hirvonen2015classification}, seems feasible for a small number of classes and directions. 
Otherwise, such a classifier would require \emph{no. of DoA classes} $\times$ \emph{no. of event classes} outputs. With only a small number of them being positive at each frame, its training would face the issue of an imbalanced dataset.
Additionally, training such a large number of classes requires an impractically huge dataset with enough examples for each class. On the other hand, the format of one DoA-regression-output per sound event class does not suffer from those limitations, but it is unable to detect multiple instances of the same class being active at different directions.

Finally, some of the submissions aimed for a parametric DoA estimation instead of a trainable DNN model \cite{Xue2019_report, Nguyen2019_report, Tan2019_report, PerezLopez2019}, including the third best system of Xue et al. \cite{Xue2019_report}. Parametric DoA estimation has the advantage that it does not require training and that it is possible to generalize to completely unseen environments, since it requires only knowledge of the directional array response. Moreover, Nguyen et al. \cite{Nguyen2019_report} had one of the smallest DoA errors in the challenge.  However, it can be more susceptible to reverberation than DNN approaches if not accompanied with additional processing, such as detection of single-source dominated time-frequency blocks \cite{Nguyen2019_report}. Interestingly,  Xue et al. \cite{Xue2019_report} did not utilize the provided theoretical steering vectors of the spatial format, but estimated them directly from the data.

\begin{table*} 
\centering
\caption{Evaluation of DCASE 2019 submissions using the joint metrics calculated in one second segments. Best system per team, in order of the official challenge ranking.}
\label{tab:seld-loc}
\begin{tabular}{c l | c c c | c c c |c c}
\toprule

\begin{tabular}[c]{@{}c@{}}Official\\rank\end{tabular} & System & LE$_{CD}$ & LR$_{CD}$ & Rank & ER$_{10^\circ}$ & F$_{10^\circ}$ & Rank & ER$_{30^\circ}$ & F$_{30^\circ}$ \\
\midrule 
{1} &
  {Kapka\_SRPOL\_2~\cite{Kapka2019}} &
  \textbf{{3.5}} $\pm$ {0.7} &
  {93.5} $\pm$ {0.9} &
  1 &
  \textbf{{0.20}} $\pm$ {0.02} &
  \textbf{{83.8}} $\pm$ {1.9} &
  1 &
  {0.13} $\pm$ {0.02} &
  {91.0} $\pm$ {1.4} 
  \\
{2} &
  {Cao\_Surrey\_4~\cite{Cao2019}} &
  {5.5} $\pm$ {0.8} &
  {94.8} $\pm$ {0.5} &
  2 &
  {0.26} $\pm$ {0.03} &
  {77.7} $\pm$ {2.5} &
  3 &
  \textbf{{0.13}} $\pm$ {0.01} &
  \textbf{{91.0}} $\pm$ {1.0}
  \\
{3} &
  {Xue\_JDAI\_1~\cite{Xue2019_report}} &
  {10.5} $\pm$ {1.5} &
  {95.4} $\pm$ {0.6} &
  5 &
  {0.30} $\pm$ {0.02} &
  {73.2} $\pm$ {2.1} &
  6 &
  {0.16} $\pm$ {0.02} &
  {87.2} $\pm$ {1.8}
  \\
{4} &
  {He\_THU\_2~\cite{He2019_report}} &
  {22.9} $\pm$ {1.6} &
  \textbf{{95.5}} $\pm$ {0.5} &
  8 &
  {0.72} $\pm$ {0.03} &
  {30.1} $\pm$ {2.8} &
  16 &
  {0.28} $\pm$ {0.03} &
  {74.6} $\pm$ {2.8}
  \\
{5} &
  {Jee\_NTU\_1~\cite{Pratik2019}} &
  {4.3} $\pm$ {0.6} &
  {93.2} $\pm$ {0.6} &
  3 &
  {0.24} $\pm$ {0.02} &
  {80.7} $\pm$ {1.8} &
  2 &
  {0.15} $\pm$ {0.01} &
  {90.9} $\pm$ {0.8} 
  \\
{6} &
  {Nguyen\_NTU\_3~\cite{Nguyen2019_report}} &
  {14.6} $\pm$ {1.6} &
  {92.1} $\pm$ {0.8} &
  9 &
  {0.51} $\pm$ {0.03} &
  {53.1} $\pm$ {3.2} &
  13 &
  {0.25} $\pm$ {0.03} &
  {80.4} $\pm$ {2.4} 
  \\
{7} &
  {MazzonYasuda\_NTT\_3~\cite{MazzonYasuda2019_report}} &
  {6.6} $\pm$ {1.0} &
  {93.4} $\pm$ {0.5} &
  4 &
  {0.30} $\pm$ {0.03} &
  {74.6} $\pm$ {2.8} &
  5 &
  {0.17} $\pm$ {0.01} &
  {88.2} $\pm$ {1.2} 
  \\
{8} &
  {Chang\_HYU\_3~\cite{Chang2019_report}} &
  {15.9} $\pm$ {2.2} &
  {90.8} $\pm$ {0.6} &
  13 &
  {0.43} $\pm$ {0.04} &
  {62.3} $\pm$ {3.7} &
  10 &
  {0.32} $\pm$ {0.03} &
  {74.4} $\pm$ {2.6}
  \\
{9} &
  {Ranjan\_NTU\_3~\cite{Ranjan2019}} &
  {14.3} $\pm$ {2.0} &
  {89.2} $\pm$ {1.0} &
  11 &
  {0.44} $\pm$ {0.04} &
  {63.1} $\pm$ {3.7} &
  10 &
  {0.31} $\pm$ {0.03} &
  {76.7} $\pm$ {2.7} 
  \\
{10} &
  {Park\_ETRI\_1~\cite{Park2019a}} &
  {6.0} $\pm$ {0.9} &
  {91.1} $\pm$ {0.6} &
  6 &
  {0.30} $\pm$ {0.02} &
  {76.2} $\pm$ {2.3} &
  4 &
  {0.20} $\pm$ {0.01} &
  {86.9} $\pm$ {1.1} 
  \\
{11} &
  {Leung\_DBS\_2~\cite{Leung2019_report}} &
  {31.4} $\pm$ {1.6} &
  {92.3} $\pm$ {0.7} &
  15 &
  {0.84} $\pm$ {0.02} &
  {17.7} $\pm$ {1.7} &
  18 &
  {0.43} $\pm$ {0.03} &
  {59.9} $\pm$ {2.6} 
  \\
{12} &
  {Grondin\_MIT\_1~\cite{Grondin2019}} &
  {8.0} $\pm$ {0.8} &
  {91.6} $\pm$ {0.8} &
  7 &
  {0.40} $\pm$ {0.03} &
  {65.4} $\pm$ {3.1} &
  9 &
  {0.19} $\pm$ {0.02} &
  {88.0} $\pm$ {1.3} 
 \\
{13} &
  {ZhaoLu\_UESTC\_1~\cite{ZhaoLu2019_report}} &
  {7.3} $\pm$ {1.0} &
  {88.3} $\pm$ {0.9} &
  10 &
  {0.39} $\pm$ {0.03} &
  {67.5} $\pm$ {3.1} &
  8 &
  {0.24} $\pm$ {0.02} &
  {83.8} $\pm$ {1.5} 
 \\
{14} &
  {Rough\_EMED\_2~\cite{Rough2019_report}} &
  {9.7} $\pm$ {1.0} &
  {88.7} $\pm$ {0.8} &
  11 &
  {0.50} $\pm$ {0.03} &
  {55.3} $\pm$ {2.8} &
  12 &
  {0.24} $\pm$ {0.02} &
  {83.4} $\pm$ {1.5} 
\\
{15} &
  {Tan\_NTU\_1~\cite{Tan2019_report}} &
  {19.0} $\pm$ {1.8} &
  {88.8} $\pm$ {1.0} &
  16 &
  {0.63} $\pm$ {0.02} &
  {41.4} $\pm$ {2.3} &
  14 &
  {0.31} $\pm$ {0.03} &
  {76.0} $\pm$ {2.4} 
\\
{16} &
  {Cordourier\_IL\_2~\cite{CordourierMaruri2019}} &
  {22.6} $\pm$ {1.4} &
  {85.8} $\pm$ {0.9} &
  17 &
  {0.78} $\pm$ {0.03} &
  {25.7} $\pm$ {2.6} &
  17 &
  {0.39} $\pm$ {0.02} &
  {67.3} $\pm$ {2.3} 
 \\
{17} &
  {Krause\_AGH\_4~\cite{Krause2019_report}} &
  {36.9} $\pm$ {1.4} &
  {86.1} $\pm$ {1.0} &
  19 &
  {0.95} $\pm$ {0.01} &
  {8.3} $\pm$ {0.8} &
  21 &
  {0.56} $\pm$ {0.02} &
  {49.5} $\pm$ {2.3}
 \\  \rowcolor{Gray}
{18} &
  {Adavanne\_TAU\_FOA~\cite{adavanne2019multi}} &
  {29.7} $\pm$ {1.3} &
  {83.8} $\pm$ {0.9} &
  18 &
  {0.95} $\pm$ {0.01} &
  {10.5} $\pm$ {1.1} &
  20 &
  {0.53} $\pm$ {0.02} &
  {56.5} $\pm$ {2.2} 
 \\
{19} &
  {Perezlopez\_UPF\_1~\cite{PerezLopez2019}} &
  {5.9} $\pm$ {0.4} &
  {81.2} $\pm$ {1.6} &
  14 &
  {0.38} $\pm$ {0.03} &
  {73.8} $\pm$ {1.8} &
  7 &
  {0.32} $\pm$ {0.03} &
  {80.2} $\pm$ {1.8} 
 \\
{20} &
  {Chytas\_UTH\_1~\cite{Chytas2019}} &
  {19.2} $\pm$ {1.5} &
  {81.0} $\pm$ {1.0} &
  19 &
  {0.70} $\pm$ {0.02} &
  {37.0} $\pm$ {2.5} &
  15 &
  {0.43} $\pm$ {0.02} &
  {67.7} $\pm$ {2.8} 
 \\
{21} &
  {Anemueller\_UOL\_3~\cite{Anemueller2019_report}} &
  {34.5} $\pm$ {1.4} &
  {82.6} $\pm$ {1.2} &
  21 &
  {0.97} $\pm$ {0.01} &
  {7.9} $\pm$ {0.9} &
  22 &
  {0.60} $\pm$ {0.03} &
  {46.8} $\pm$ {2.3} 
 \\
{22} &
  {Kong\_SURREY\_1~\cite{Kong2019_report}} &
  {42.7} $\pm$ {2.1} &
  {82.2} $\pm$ {1.0} &
  22 &
  {0.92} $\pm$ {0.01} &
  {11.0} $\pm$ {1.4} &
  19 &
  {0.65} $\pm$ {0.02} &
  {41.3} $\pm$ {2.5} 
 \\
{23} &
  {Lin\_YYZN\_1~\cite{Lin2019_report}} &
  {92.7} $\pm$ {20.9} &
  {1.1} $\pm$ {0.4} &
  23 &
  {1.04} $\pm$ {0.01} &
  {0.0} $\pm$ {0.0} &
  23 &
  {1.04} $\pm$ {0.01} &
  {0.2} $\pm$ {0.2} 
\\
\bottomrule
\end{tabular}
\end{table*}

\begin{figure*}[!t]
  \centering
  \subfloat[Location-aware detection results]{
  \includegraphics[width=0.5\textwidth]{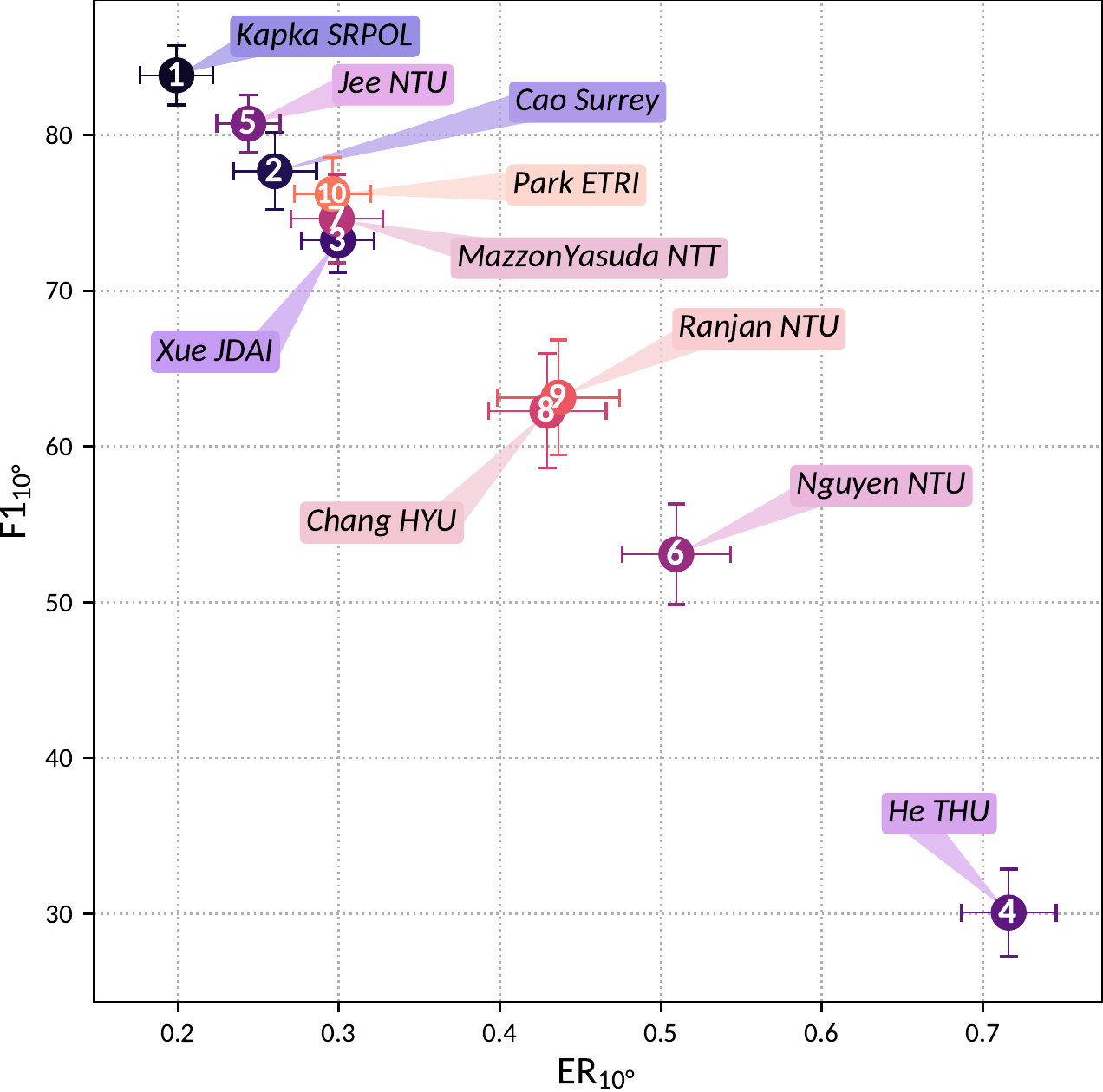} 
  \label{fig:CI-SELDdet}}  
  \subfloat[Class-aware localization results]{
  \includegraphics[width=0.5\textwidth]{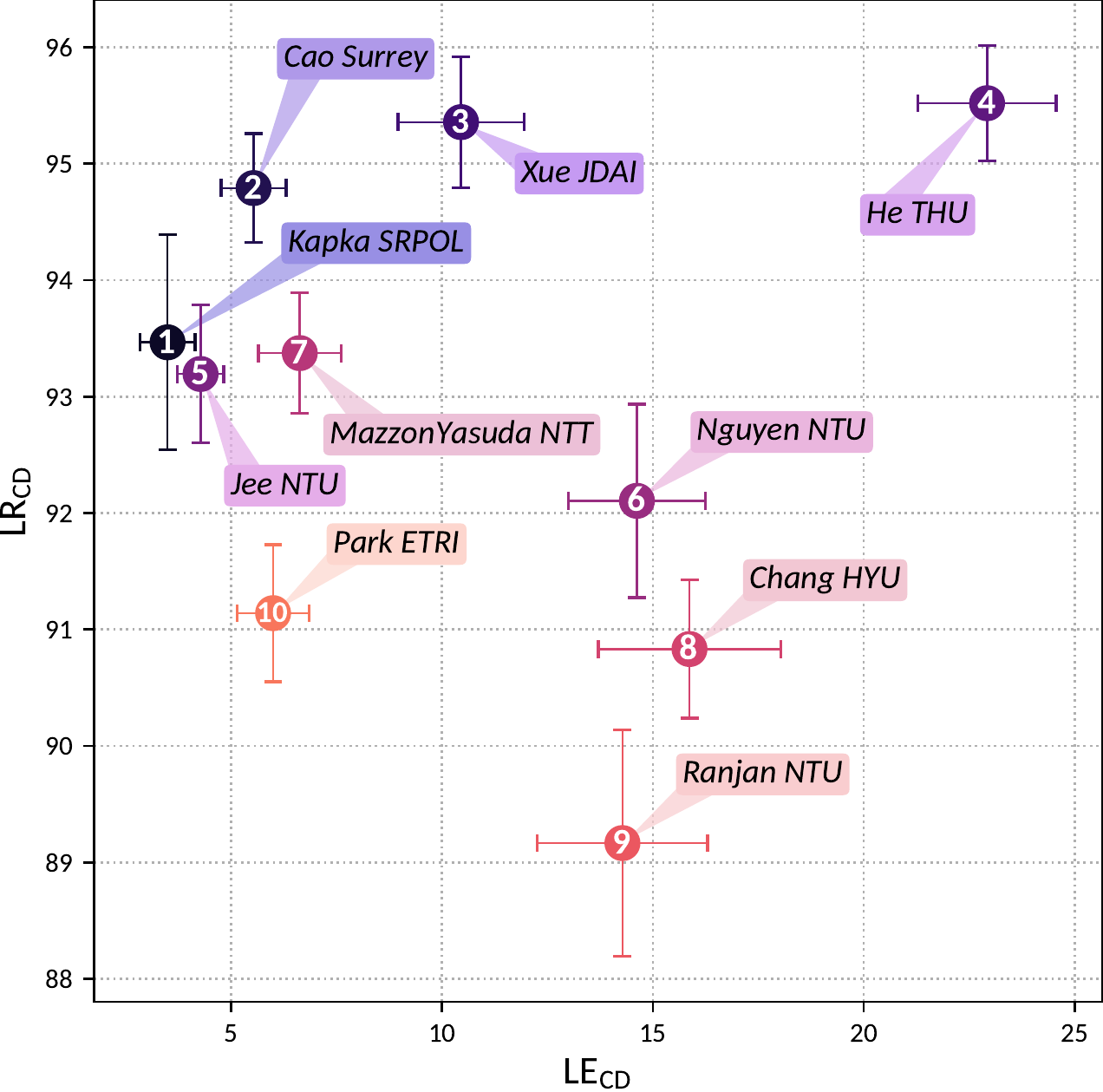} 
  \label{fig:CI-SELDloc}}
\caption{Joint detection and localization performance of top 10 systems (best system per team). The official rank of the systems is indicated in the center of the marker for each scatter plot.}
\label{fig:seld-metrics}
\vspace{-8pt}
\end{figure*}

\section{Reevaluation of challenge entries using joint metrics}
\label{sec:reeval}

We evaluate all the systems submitted to DCASE 2019 Challenge Task 2 using the proposed joint measures in order to determine the most suitable single metric that encompasses all aspects when representing system performance in a single number. We compute all metrics in one-second segments and evaluate the location-aware detection metrics with an angular error threshold of 10 and 30 degrees. 
The results are presented in Table \ref{tab:seld-loc}, in order of the official challenge rank. Confidence intervals for all metrics were calculated according to the jackknife procedure by leaving out one file at a time for the partial evaluation. New cumulative ranks are estimated similar to the official ranks based on the proposed joint measures for the purpose of system comparison. The top 10 systems from Table \ref{tab:seld-loc} are also presented in Fig. \ref{fig:seld-metrics}.

\subsection{Analysis of systems}

The independent localization and evaluation metrics ($ECR,LE,F1,ER$) are more permissive than the joint ones ($LR_{CD}, LE_{CD}, F_{10^\circ}, ER_{10^\circ}$). We chose a threshold of $10^\circ$ for a relatively strict localization criteria with respect to the average localization error of the systems presented in Table~\ref{tab:official-results-sys-char}. A ranking based on the new metrics is expected to be different at least for some of the submissions. Table \ref{tab:seld-loc} presents new ranks computed between class-dependent localization ($LR_{CD}, LE_{CD}$) and location-dependent classification ($F_{10^\circ}, ER_{10^\circ}$). Systems with equal ranks indicate that the sum of the individual ranks for each pair of metrics was the same. The greatest changes on the top ten systems seem to be induced by the location-dependent classification ($F_{10^\circ}, ER_{10^\circ}$), which is to be expected since it penalizes inadequately localized detections with a strict threshold of $10^\circ$. 

In general, it can be observed that submissions which employed separate localization and detection systems and did not handle association of the two properly were likely to slip in their ranks. This is especially evident on the systems that assigned randomly DoAs to detections, such as \textit{Nguyen NTU}~\cite{Nguyen2019_report} and \textit{Ranjan NTU}~\cite{Ranjan2019}, including the best localization method of \textit{Chang HYU}~\cite{Chang2019_report}. Their association problems are revealed both by their large drop in detection scores ($F_{10^\circ}, ER_{10^\circ}$) and by the large error increase between their original $LE$ and the class-dependent one $LE_{CD}$.

Methods that performed significantly better detection than localization, such as \textit{Xue JDAI}~\cite{Xue2019_report}, \textit{He THU}~\cite{He2019_report}, and \textit{Leung DBS}~\cite{Leung2019_report} also slipped in their ranks. This is mostly due to three of the original metrics ($F1, ER, ECR$) being directly associated to detection performance, boosting their overall rank. This imbalance is diminished with the new metrics, resulting in the drop of the aforementioned systems.

Among the methods that performed proper data association, the ones who had better localization scores~\cite{PerezLopez2019,Grondin2019,Park2019a,ZhaoLu2019_report,MazzonYasuda2019_report,Chytas2019} and not the best detection scores improved in their ranks, due to the detection bias of official rankings mentioned above. Two examples worth mentioning are \cite{Park2019a,PerezLopez2019}. The multi-task training strategy of \textit{Park ETRI}~\cite{Park2019a} showed its benefits when evaluated jointly, taking them to 4th place. \textit{PerezLopez UPF}~\cite{PerezLopez2019} leaped from 19th place below the baseline to 7th place. Both systems achieved such rank advances when evaluated with the strict location-dependent detection ($F_{10^\circ}, ER_{10^\circ}$).

Even though the rank for the more permissive $30^\circ$ location-dependent detection metrics ($F_{30^\circ}, ER_{30^\circ}$) is not displayed in Table \ref{tab:seld-loc}, it is closer to the original challenge ranking. 
This is explained a) by the more relaxed threshold, which as it increases it causes the metrics to approach their independent detection counterparts. B) by the fact that the threshold is larger than the average $LE_{CD}$ of about $20^\circ$ between systems.

\subsection{Metrics analysis}

\begin{figure*}
    \centering
    \includegraphics[width=\textwidth]{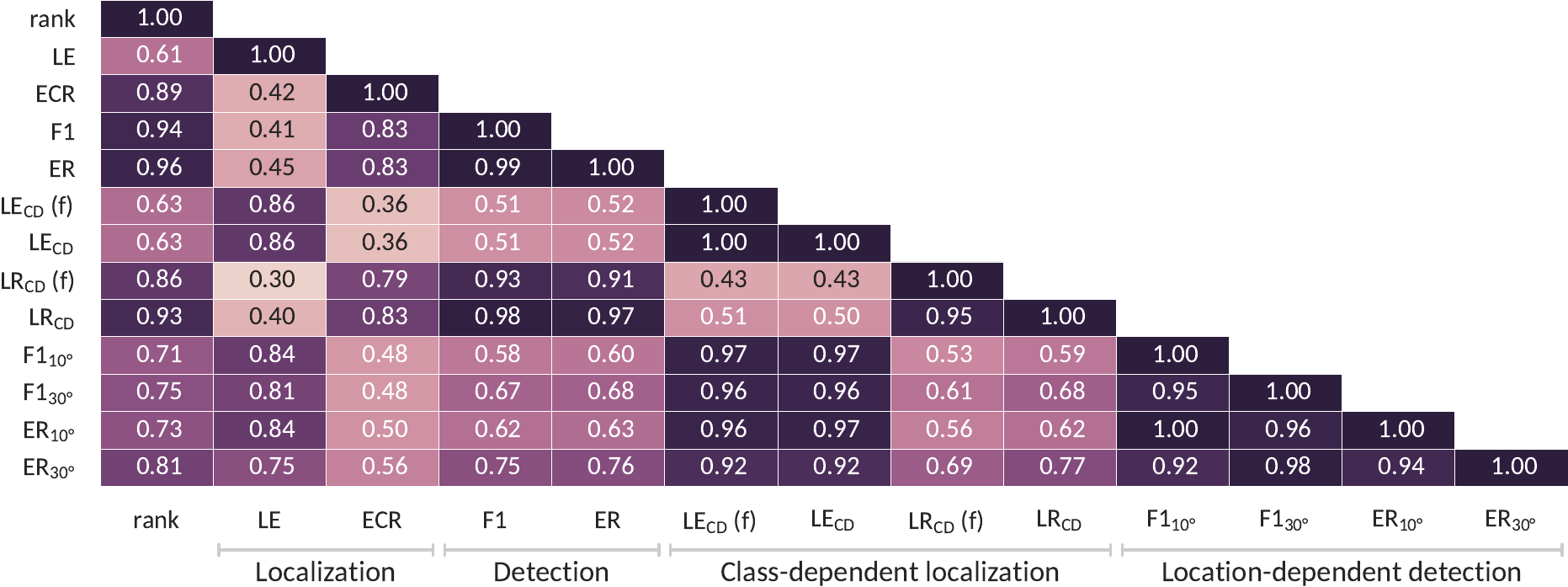}
    \caption{Correlation between ranking order of submissions according to the different metrics and the official ranking in the challenge.}
    \label{fig:correlation}
\end{figure*}

The analysis of the metrics was performed using Spearman's rank correlation coefficient \cite{spearman1904proof} to calculate how they correlate to each other. The correlation was calculated between all pairs of considered metrics, using the evaluated performance of all submissions to the task. Our purpose is to determine which single metric is capable of representing the desired properties of the system in terms of localization and detection, instead of using the compound of four separate metrics as done in the challenge ranking.
We rank all submissions using each metric separately and evaluate how correlated the different rankings are. Correlation values are presented in Fig. \ref{fig:correlation}. The metrics marked with ($f$) are calculated frame-wise (in this case 20 ms).  
Among the four individual metrics ($LE$, $ECR$, $F1$, and $ER$), the detection scores ($F1$ and $ER$) are highly correlated with the ranking, indicating that good detection performance was important for obtaining a top rank. The localization error is less correlated with the overall rank.

Among the joint metrics, the class-dependent $LR_{CD}$ score is highly correlated with the official ranking, more so for the segment-based than the frame-based measurement. 
This behavior is noticed in all metrics, with the more permissive metric being more correlated to the overall rank: a) segment-based $LR_{CD}$ is more correlated to the rank than frame-based $LR_{CD} (f)$ and b) metrics with $30^\circ$ threshold are more correlated to the rank than metrics with $10^\circ$ threshold. This can be explained by the fact that joint metrics first perform the data association between detected and localized sound sources. The more permissive metrics allow a higher proportion of matches, which in turn is closer to the matching done by the detection-only and separation-only metrics.

We observe similar behavior between metric pairs with and without data association: a) correlation between localization-only metrics $LE$ and $ECR$ is moderate and similar to the one between $LE_{CD}$ and $LR_{CD}$. b) High correlation is observed between detection-only $ER$ and $F1$; similarly for the corresponding data associated versions. On the other hand, the correlation between detection-only $ER$ and its counterparts  $ER_{10^\circ}$ or $ER_{30^\circ}$ is moderate. Similar behaviour is observed between $F1$ and its counterparts $F1_{10^\circ}$ or $F1_{30^\circ}$. Basically, the data association makes the metrics less permissive (in a similar manner as the higher correlation for the more permissive threshold of $30^\circ$ than for $10^\circ$). 

Among the proposed joint-metrics, $LR_{CD}$ has the best correlation (0.93) with the official DCASE2019 rankings, that is presumed to be a good approximation of the overall system performance. However, $LE_{CD}$ is only moderately correlated (0.50) with $LR_{CD}$, hence, selecting an SELD model based on just $LR_{CD}$ might not always guarantee the best $LE_{CD}$. On the other hand, the location-aware detection metrics are highly correlated with each other ($ER_{10^\circ}$ vs. $F1_{10^\circ}$ or $ER_{30^\circ}$ vs. $F1_{30^\circ}$) and have moderately high (0.71-0.81) correlation with the official rank. 
Furthermore, for a given distance threshold, the error rate metrics are more correlated to the official rank than the F1-scores and they are also highly correlated with $LE_{CD}$. 
If choosing a SELD model has to be limited to a single metric only, it seems that the error rate ($ER_{10^\circ}$/$ER_{30^\circ}$) is a suitable choice, since it combines high correlation with the original ranking, with the ranking based on ($F1_{10^\circ}$/$F1_{30^\circ}$), as well as the localization ranking of $LE_{CD}$. Hence, it is expected to guarantee an overall good SELD performance, a good counterpart F1-score and a low localization error.

\subsection{Discussion}

The very high performance of the top ranked systems, of a few degrees of mean localization error and more than 83$\%$ F1 score in the stricter setting, reveals additionally that the state of the art can potentially handle more challenging conditions than those reflected on the current dataset.
The simulated spatial recordings, even though acoustically realistic, contained only static events well separated between them by at least 10$^\circ$. Furthermore, the room IRs were captured in large open spaces and at fairly close distances from the microphone resulting in high direct-to-reverberant ratios, while the ambient noise was added at a very high SNR. As a consequence, the spatial and spectral characteristics of the events were not significantly corrupted by them and the methods had to learn mostly a model of the directional array response to infer location.
Such conditions of up to two simultaneous foreground sound events of interest at differing directions and at 1--2 m away from the listener, in the presence of reverberation and low background noise can still occur frequently in real-life, but they are, of course, only a subset of real spatial sound scenes and of the associated challenges for SELD systems.
Most of these considerations were addressed in the recent dataset for the new DCASE2020 challenge \cite{Politis2020_task3_report}. A significant advance is the introduction of reverberant moving sources, still based on captured RIRs from real spaces \cite{adavanne2019localization, Politis2020_task3_report}. Moreover, ambient noise occurs at varying levels, reverberant conditions are stronger and more varied, and event locations do not occur in a sparse regular grid but can vary more or less continuously. Hence, after DCASE2019 confirmed that informed engineering can solve the SELD task successfully under the restricted conditions of its dataset, the DCASE2020 challenge focuses on presenting more challenging evaluation conditions closer to reality.

It should be pointed out that a rigorous analysis of the metrics and of their accuracy, consistency, and behaviour in general is still open and remains a topic for future work. It should be also emphasized that until such work has been done, the proposed results should be seen as empirical; they are based on observation and on correlating the proposed performance measurements with the expected behaviour of the measured systems. Similar empirical results were obtained in a second setting in \cite{Mesaros_2019_WASPAA}, where a learning-based SELD system was measured while it was trained, with all scores increasing in accordance with its optimization objective.

\section{Conclusions and future work}
\label{sec:concl}
This work presented and analyzed the submissions of the DCASE2019 SELD challenge, with a discussion on general and individual characteristics of the systems, how those reflected on their performance, and a comprehensive evaluation. This first challenge revealed a strong community focused on the joint localization and detection, coming both from the audio machine learning and the array signal processing fields. Compared to the few related studies before the challenge, the participants demonstrated strong advances in terms of SELD modeling, engineering, and in terms of raw performance. The majority of submissions surpassed the baseline with a large margin and the best ones reached almost perfect localization and detection scores.

Taking into account the advances in the recent DCASE2020 SELD challenge, we can envision some of the challenges in a SELD task that have not been addressed yet. In terms of the spatial properties of the scene, two points not addressed yet are moving receivers (together with moving sources) and directional interferences which represent clearly localized sounds of unknown types. Both of these properties are expected to be introduced in the upcoming challenges, after DCASE2020.
Beyond spatial characteristics, an evolution of the challenge and its datasets would consider the overall spatiotemporal scene consistency. At the moment events are randomly chosen and spatialized. A realistic scene generator should spatialize events that fit a given space at their most probable locations, while respecting real-life co-occurence probabilities. Such consistency between space, sound source locations, respective sound emitting actions, and the sound events associated with all the above remains a topic for future research.

\bibliographystyle{IEEEtran}
\bibliography{references.bib}

\end{document}